\documentclass{svproc}

\usepackage{url}

\usepackage{graphicx}

\usepackage{booktabs}
\usepackage{tabularx}
\usepackage{multirow}
\usepackage{array}

\usepackage{amsmath}
\usepackage{amssymb}

\usepackage{algorithm}
\usepackage{algpseudocode}

\usepackage{csquotes}

\usepackage[table]{xcolor}
\usepackage{hyperref}
\hypersetup{hidelinks}

\usepackage{orcidlink}
\providecommand{\orcidID}[1]{\,\orcidlink{#1}}

\usepackage[acronym]{glossaries}
\glsdisablehyper
\newacronym{ci}{CI}{Computational Intelligence}
\newacronym{ser}{SER}{Speech Emotion Recognition}
\newacronym{asr}{ASR}{Automatic Speech Recognition}
\newacronym{llm}{LLM}{Large Language Model}
\newacronym{mfcc}{MFCC}{Mel-Frequency Cepstral Coefficient}
\newacronym{rtf}{RTF}{Real-Time Factor}
\newacronym{esn}{ESN}{Echo State Network}
\newacronym{dpia}{DPIA}{Data Protection Impact Assessment}

\graphicspath{{figures/}}

\begin{document}
\mainmatter

\title{EmotionAI: A Privacy-Preserving Computational Intelligence Pipeline for Speech-Emotion-Grounded Conversational Analysis}
\titlerunning{Privacy-Preserving CI for Emotion-Grounded Conversational Analysis}
\author{Wai Laam Mak\orcidID{0009-0004-8275-8567}\textsuperscript{*} \and
Isibor Kennedy Ihianle\orcidID{0000-0001-7445-8573} \and
Pedro Machado\orcidID{0000-0003-1760-3871}\textsuperscript{*}}
\authorrunning{W.~L. Mak \and I.~K. Ihianle \and P. Machado}
\tocauthor{Wai Laam Mak, Isibor Kennedy Ihianle, Pedro Machado}
\institute{School of Science and Technology, Nottingham Trent University, Nottingham, UK\\
\email{N1183564@my.ntu.ac.uk; isibor.ihianle@ntu.ac.uk; pedro.machado@ntu.ac.uk}\\
\textsuperscript{*}\,Corresponding authors}

\maketitle

\begin{abstract}
Reviewing recorded interviews for affective cues such as composure and agitation is slow and subjective, and cloud services that could automate the task require sensitive audio to leave the device.
EmotionAI is a fully local \gls{ci} pipeline that couples \gls{ser} with generative reasoning.
Speaker diarisation, Whisper \gls{asr} and a wav2vec2 emotion classifier produce per-segment affective evidence, and an adversarial three-model local \gls{llm} panel turns that evidence into timestamp-grounded, citation-constrained answers.
Zero-shot evaluation on the RAVDESS four-class English subset (n\,=\,672) measures the cost of cross-corpus transfer: the deployed classifier scores 48.8\% accuracy, above random (24.9\%) and majority (28.6\%) baselines but below an in-domain MFCC~+ logistic-regression comparator (71.0\%).
The complete pipeline runs in a mean 157\,s on CPU (real-time factor $\approx$1.33) with zero external calls.
The contribution is not state-of-the-art SER but an auditable, privacy-preserving integration of imperfect affective evidence into grounded conversational analysis.
\end{abstract}

\section{Introduction}

Interviews, panels and question-and-answer sessions carry affective information in the voice that the transcript alone discards \cite{tenbosch2003}, and reviewers re-listen to recordings to recover that context \cite{li2024slt}, making manual review slow, subjective and hard to audit.
Cloud multimodal services that could automate the task upload audio to third-party infrastructure, unacceptable in many journalistic, clinical and academic settings \cite{aloufi2021,backstrom2025}.
A tool is therefore needed that extracts, visualises and reasons over affective content alongside the transcript while keeping all processing on the user's own machine, and because no single model can classify affect, aggregate it over time and explain it in language, the requirement is \gls{ci} integration rather than any one classifier.
The language of \enquote{credibility} or \enquote{truthfulness} is deliberately avoided: the system supports \emph{affective review} and \emph{speaker-state analysis}, and affect inference is explicitly \emph{not} deception detection (Section~\ref{sec:ethics}).

The work contributes an audio-emotion-to-language pipeline fusing diarisation, ASR and wav2vec2 SER with an adversarial three-model local \gls{llm} panel served via Ollama, a question-answering mode whose prompt forces the generator to cite segment timestamps and emotion evidence, and an empirical evaluation of cross-corpus SER robustness and local CPU runtime given an imperfect classifier.

Three evaluation questions organise the results: \textbf{EQ1 (SER robustness)} how the pre-trained SER model transfers zero-shot relative to in-domain and classical baselines, \textbf{EQ2 (Q\&A utility)} whether emotion grounding yields more traceable answers than transcript-only prompting, and \textbf{EQ3 (deployment feasibility)} whether the pipeline runs on commodity CPU hardware, and how fast.

\section{Related Work}

\paragraph{Speech emotion recognition.}
Classical SER pairs hand-crafted spectral descriptors, typically \glspl{mfcc}, with a shallow classifier \cite{oshaughnessy2025}, whereas self-supervised models such as wav2vec2 learn latent representations directly from raw waveforms without manual feature engineering \cite{baevski2020}.
Laboratory accuracy does not transfer across corpora: taxonomies and acoustic realisations differ between IEMOCAP and RAVDESS, and zero-shot transfer degrades sharply \cite{lashkarashvili2024}.
An MFCC\,+\,logistic-regression baseline and an \gls{esn} \cite{ibrahim2022} separate hand-crafted features and temporal dynamics from pre-trained transformer embeddings.

\paragraph{Local and multimodal LLM pipelines.}
Open-weight LLMs (Llama~3, Qwen2.5, Gemma~3) now run capably on consumer hardware \cite{gemma2025,grattafiori2024,qwen2025} but are text-only: a confident utterance cannot be distinguished from a hesitant one when the words are identical.
Native audio LLMs such as Qwen2-Audio and SALMONN ingest waveforms directly, but they bind classifier and generator into a single network that cannot be split into the conditions EQ2 compares, and their footprint exceeds the commodity-CPU budget of EQ3.
Structured emotion metadata can instead be injected into a text LLM as prompt context \cite{lin2024,thimonier2025}, keeping classifier and generator independently replaceable, the pattern adopted here.

Explainability for interview review means tracing each assertion to a specific segment and emotion score, and Russell's valence--arousal circumplex \cite{russell1980} grounds the indicators in psychological theory.
On-device processing removes the dominant risk in handling biometric-adjacent audio \cite{aloufi2021,backstrom2025}, while emotion recognition carries bias and dual-use risks demanding advisory-only framing \cite{katirai2024,sedenberg2017}.
Few published systems close the loop between a locally deployed emotion classifier and a locally hosted LLM in one pipeline.

\section{Method}
\label{sec:method}

\begin{figure}[htbp]
\centering
\includegraphics[width=0.56\linewidth]{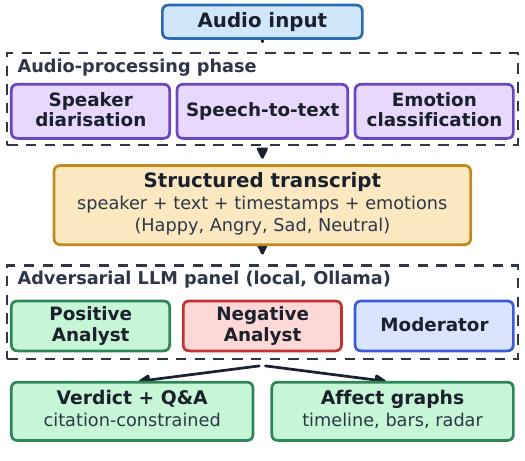}
\caption{System architecture. The audio phase (top) runs pyannote~3.1 diarisation with WeSpeaker
verification, Whisper-medium ASR and wav2vec2 emotion classification; a cleanup step then frees
all audio models before the LLM phase (bottom) loads Llama~3.2:3B, Qwen~2.5:3B and Gemma~3:4B in
sequence, bounding peak memory to the largest single model. All processing is on the local host.}
\label{fig:arch}
\end{figure}

EmotionAI runs two sequential phases (Fig.~\ref{fig:arch}): the audio phase diarises and verifies speakers, transcribes each turn and classifies per-segment emotion; the LLM phase then loads, runs and unloads the three analysts strictly in sequence (\texttt{keep\_alive=0}).
A cleanup step between the phases frees every audio-stage model and forces garbage collection before any LLM loads, bounding peak memory by the largest single model.

\begin{table}[htbp]
\centering
\footnotesize
\caption{The six-stage EmotionAI pipeline. Stages communicate only through the listed outputs.}
\label{tab:method}
\renewcommand{\arraystretch}{1.05}
\begin{tabularx}{\linewidth}{>{\raggedright\arraybackslash}p{0.15\linewidth}XX}
\toprule
\textbf{Stage} & \textbf{Library / model} & \textbf{Output handed to next stage}\\
\midrule
1. Ingestion & librosa (resample to 16\,kHz) & 16\,kHz mono float32 waveform\\
2. Diarisation & pyannote 3.1 + WeSpeaker & (start, end, speaker) turns\\
3. ASR & Whisper medium & per-turn \{speaker, start, end, text\} with word-level timestamps\\
4. Emotion & wav2vec2-large SER & per-segment 4-class softmax \{Hap, Ang, Sad, Neu\}\\
5. Assembly & internal & JSON transcript + structured emotion-annotated prompt\\
6. LLM panel & Ollama local LLM panel & per-speaker positive / negative / moderator verdict\\
\bottomrule
\end{tabularx}
\end{table}

Table~\ref{tab:method} summarises the six stages, which build on pyannote 3.1, Whisper medium and wav2vec2-large-superb-er \cite{baevski2020,bredin2019,radford2022,wang2022}.
The WeSpeaker override threshold (cosine similarity 0.12) was tuned on recordings with ground-truth speaker labels.
Fixed output schemas at every stage boundary keep components independently replaceable and isolate malformed outputs to their own stage.

\begin{figure}[htbp]
\centering
\includegraphics[width=0.56\linewidth]{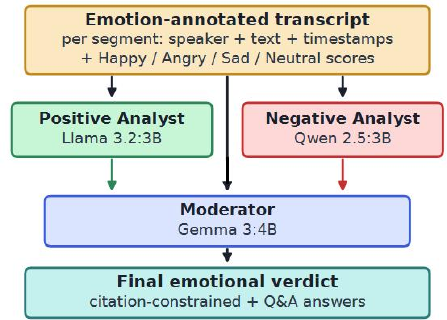}
\caption{Adversarial three-model panel and prompt-assembly flow. The Positive and Negative
Analysts each read the same emotion-annotated transcript, and the Moderator reconciles their
two analyses into the citation-constrained verdict and answers. Pairing opposed briefs before
moderation is intended to reduce the interpretive bias of a lone summariser.}
\label{fig:panel}
\end{figure}

The assembled prompt pairs each diarised segment with its per-class emotion probabilities (Fig.~\ref{fig:panel}): Llama~3.2:3B as a Positive Analyst and Qwen~2.5:3B as a Negative Analyst each read it independently, and Gemma~3:4B as a Moderator reconciles their two analyses into the final verdict and answers subsequent questions against the emotion-annotated transcript.

The prompts forbid quoting raw probability strings, instead requiring the panel to translate emotion distributions into natural language and cite segment timestamps, making the Q\&A \emph{timestamp-grounded} and auditable.

The panel targets a failure mode of a lone summariser, which tends to commit to one interpretation of ambiguous evidence and state it with unwarranted confidence: opposed briefs for Llama~3.2:3B (favourable) and Qwen~2.5:3B (critical) force two readings to exist before Gemma~3:4B reconciles them.
Drawing the roles from three model families is intended to keep any one family's interpretive habits from dominating the verdict, and the 3--4\,B scale fits sequential loading; Section~\ref{sec:results} tests both the panel and the family mix.

Per-speaker indicators aggregate the segment-level emotion probabilities for each speaker, bounded to a 0--100 scale from Russell's circumplex \cite{russell1980}.
For a speaker $u$ with segment set $S_u$ and per-segment softmax $\mathbf{p}_s=(p^{\mathrm{hap}}_s, p^{\mathrm{ang}}_s, p^{\mathrm{sad}}_s, p^{\mathrm{neu}}_s)$, the mean affect vector $\bar{\mathbf{p}}_u$ and the Composure indicator $C_u$ are
\begin{equation}
\bar{\mathbf{p}}_u = \frac{1}{|S_u|}\sum_{s\in S_u}\mathbf{p}_s,
\qquad
C_u = 100\,\big(\bar{p}^{\,\mathrm{hap}}_u + \bar{p}^{\,\mathrm{neu}}_u\big)\in[0,100].
\end{equation}
Valence and Arousal are analogous affine projections of $\bar{\mathbf{p}}_u$ onto Russell's axes, likewise bounded to $[0,100]$, and Table~\ref{tab:indicators} gives the reading of each indicator.
All three are screening aids rather than measurements: an unexpected value should send the reviewer to the segments that produced it for re-listening, and no single score is a judgement of honesty or competence.
Because the indicators inherit the classifier's errors, including the weak Sad recall reported in Section~\ref{sec:results}, the interface presents them alongside the segment-level evidence rather than in isolation.

\begin{table}[htbp]
\centering
\footnotesize
\caption{Reading the per-speaker indicators. All are bounded to $[0,100]$ and traceable to the segments that produced them.}
\label{tab:indicators}
\renewcommand{\arraystretch}{1.1}
\begin{tabularx}{\linewidth}{>{\raggedright\arraybackslash}p{0.14\linewidth}XX}
\toprule
\textbf{Indicator} & \textbf{High score means} & \textbf{Low score means}\\
\midrule
Composure & speaker sounded predominantly calm or positive & anger or sadness dominated the classifier's reading\\
Valence & speech sounded pleasant overall & speech sounded negative overall\\
Arousal & activated, energetic delivery & flat, low-energy delivery\\
\bottomrule
\end{tabularx}
\end{table}

\section{Experimental Setup}
\label{sec:setup}

\begin{figure}[htbp]
\centering
\includegraphics[width=0.66\linewidth]{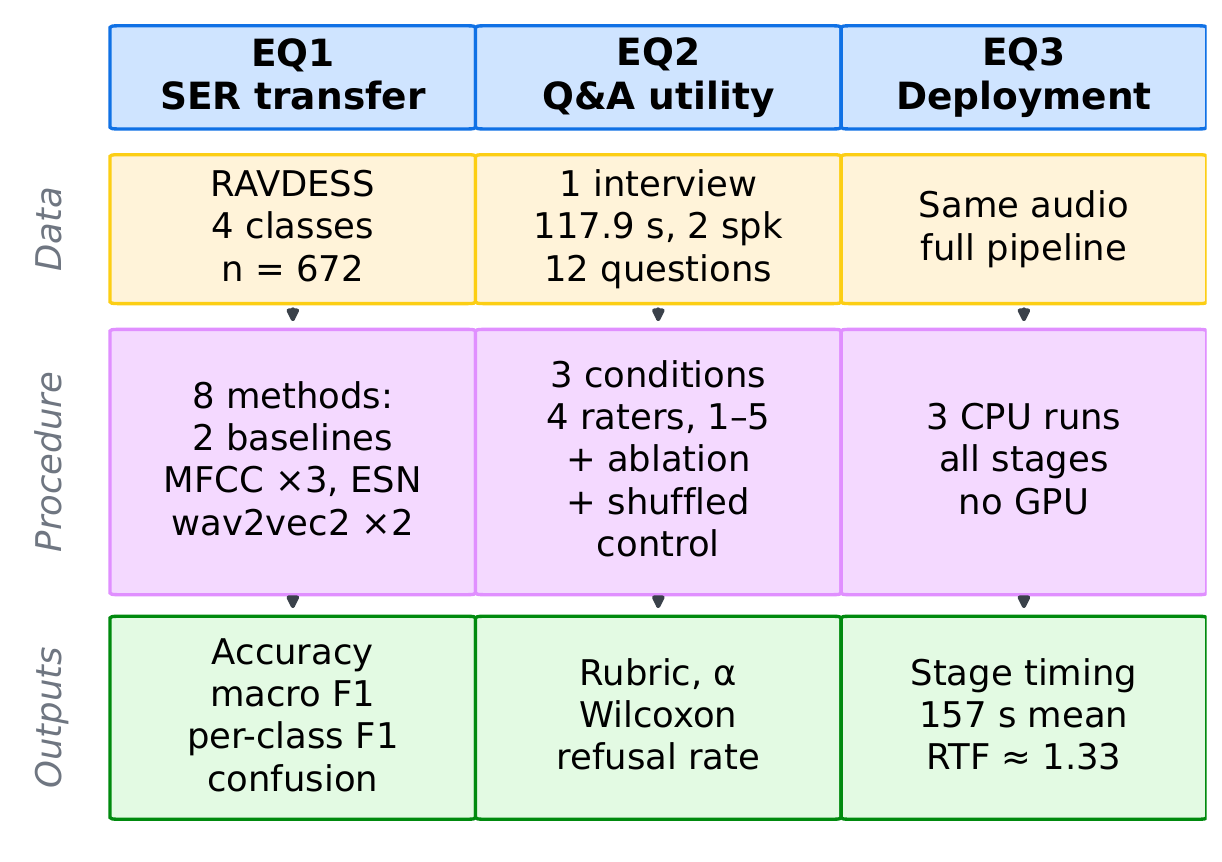}
\caption{Evaluation protocol for the three evaluation questions. EQ1 (SER robustness) compares
eight classifiers on the RAVDESS four-class subset. EQ2 (Q\&A utility) runs a twelve-question,
three-condition rubric study scored by four raters on one recorded interview, plus an
automatically scored panel ablation and corrupted-evidence control on the same questions and
recording (Table~\ref{tab:conditions}). EQ3 (deployment feasibility) benchmarks CPU runtime
over three runs. All evaluation runs on the local host.}
\label{fig:evalproto}
\end{figure}

The three evaluation questions are addressed on separate evidence, as summarised in Fig.~\ref{fig:evalproto}.

\begin{table}[htbp]
\centering
\footnotesize
\caption{The five Q\&A evaluation conditions. The first three form the four-rater rubric study; the last two, together with the emotion-informed condition rerun as \emph{panel (deployed)}, form the automatically scored panel ablation and corrupted-evidence control.}
\label{tab:conditions}
\renewcommand{\arraystretch}{1.1}
\begin{tabularx}{\linewidth}{>{\raggedright\arraybackslash}p{0.21\linewidth}p{0.19\linewidth}p{0.15\linewidth}p{0.15\linewidth}X}
\toprule
\textbf{Condition} & \textbf{Emotion evidence} & \textbf{Verdict source} & \textbf{Prompt} & \textbf{Scoring}\\
\midrule
Emotion-informed & softmax as text & panel & deployed & 4 raters + automatic, 3 repeats\\
Transcript-deployed & removed & panel & deployed & 4 raters\\
Transcript-softened & removed & panel & Rule 6 relaxed & 4 raters\\
Single-model & softmax as text & Gemma 3:4B alone & deployed & automatic, 3 repeats\\
Shuffled (control) & permuted & panel & deployed & automatic, 3 repeats\\
\bottomrule
\end{tabularx}
\end{table}

\paragraph{Dataset.}
The SER benchmark is RAVDESS \cite{livingstone2018}: 1{,}440 utterances by 24 North-American actors (12 female, 12 male) at 48\,kHz, fitting the pipeline's English interview target and the deployed classifier's English IEMOCAP fine-tuning.
Filtering to the four classes shared with the classifier's label set yields 672 utterances, unevenly distributed (Happy 192, Angry 192, Sad 192, Neutral 96) and resampled to 16\,kHz mono before inference.

Eight SER methods are compared on the same 672-clip subset: uniform-random and majority-class baselines, MFCC\,+\,$\ell_2$ logistic regression at $n_\mathrm{mfcc}\in\{13,20,40\}$ under stratified five-fold cross-validation, an ESN (500-neuron reservoir, spectral radius 0.9, leak rate 0.3, ridge readout, same protocol), and the zero-shot wav2vec2-base/large-superb-er checkpoints fine-tuned on IEMOCAP.
Accuracy, macro and weighted F1, and per-class F1 are reported with row-normalised confusion matrices.
A text-only sentiment baseline was not run; emotion-informed versus transcript-only is compared at the LLM level instead.

\paragraph{Q\&A evaluation protocol.}
EQ2 uses the five conditions of Table~\ref{tab:conditions} on 12 questions about a two-speaker English interview.
Emotion evidence, where present, is the per-segment softmax rendered as text (e.g.\ \enquote{Happy 89\%, Angry 0\%, Sad 0\%, Neutral 11\%}), a deliberate variation on the interface's compact top-class display; the softened prompt relaxes Rule~6 (the mandate to cite emotion scores) to \enquote{cite specific timestamps and quoted text as evidence}; shuffled permutes the emotion distributions across segments while transcript, speakers and timestamps stay fixed.
The rubric answers were blinded, presented in random order and scored by four raters on a five-dimension 1--5 rubric (traceability, faithfulness, timestamp accuracy, emotional specificity, groundedness); templated refusals were detected automatically, excluded from rubric scoring and reported as a condition-level rate.
The two ablation conditions and the deployed configuration run three times each, giving 108 answers scored automatically for refusals, cited timestamps, emotion terms, percentage citations and length, with no human re-rating.
A further pair of conditions isolates model diversity by regenerating the verdict with the deployed role prompts unchanged but all three roles played by one model (Gemma~3:4B) rather than three families, again over three repeats.

\paragraph{Hardware, runtime and reproducibility.}
EQ3 measures the complete pipeline entirely on CPU over the 117.9\,s two-speaker interview across three independent runs, reporting mean and min--max per stage, on an AMD Ryzen~7 7800X3D (8 cores, 64\,GB RAM, Windows~11, GPU disabled).
All models are pre-trained and publicly available, and the filtering rule, seeded protocols (random-baseline seed 42), evaluation scripts and raw per-clip predictions are available from the authors, so every quoted number regenerates from the source audio.

\section{Results}
\label{sec:results}

\paragraph{SER robustness (EQ1).}
Across the eight methods (Table~\ref{tab:res}), the in-domain MFCC\,+\,logistic-regression model peaks at $n_\mathrm{mfcc}=20$ (71.0\% accuracy, macro-F1 0.688) and the ESN reaches 68.3\% (macro-F1 0.676), while the deployed wav2vec2-large checkpoint trails at 48.8\% (macro-F1 0.394).
The decisive result is per-class collapse: wav2vec2-large scores F1\,=\,0.010 on Sad, against 0.677 for the in-domain MFCC model.
The gap reflects \emph{protocol}, not capacity: the classical models are trained on RAVDESS under cross-validation while the transformers transfer zero-shot from IEMOCAP, so cross-corpus shift is the bottleneck and light in-domain fine-tuning is the direct remedy \cite{lashkarashvili2024}.

\begin{table}[htbp]
\centering
\caption{Emotion classification on the RAVDESS four-class subset (n\,=\,672, English). MFCC and
ESN methods use stratified five-fold cross-validation; transformer methods are zero-shot.
Best values bold$^{\bigstar}$, worst marked $^{\dagger}$, shading supplementary.}
\label{tab:res}
\resizebox{\textwidth}{!}{%
\begin{tabular}{lrrrrrrrrrr}
\toprule
\textbf{Method} & \textbf{Feat.} & \textbf{Acc} & \textbf{Mac-P} & \textbf{Mac-R} & \textbf{Mac-F1} & \textbf{Wt-F1} & \textbf{F1-Hap} & \textbf{F1-Ang} & \textbf{F1-Sad} & \textbf{F1-Neu}\\
\midrule
Random (uniform) & n/a & \cellcolor{red!20}24.9\%$^{\dagger}$ & 0.249 & \cellcolor{red!20}0.243$^{\dagger}$ & 0.242 & 0.255 & 0.307 & 0.268 & 0.242 & 0.150\\
Majority-class & n/a & 28.6\% & \cellcolor{red!20}0.071$^{\dagger}$ & 0.250 & \cellcolor{red!20}0.111$^{\dagger}$ & \cellcolor{red!20}0.127$^{\dagger}$ & 0.444 & \cellcolor{red!20}0.000$^{\dagger}$ & \cellcolor{red!20}0.000$^{\dagger}$ & \cellcolor{red!20}0.000$^{\dagger}$\\
MFCC + LogReg (n=13) & 26-dim & 66.8\% & 0.650 & 0.643 & 0.646 & 0.666 & 0.665 & 0.818 & 0.599 & 0.503\\
MFCC + LogReg (n=20) & 40-dim & \cellcolor{green!25}\textbf{71.0\%}$^{\bigstar}$ & \cellcolor{green!25}\textbf{0.690}$^{\bigstar}$ & \cellcolor{green!25}\textbf{0.686}$^{\bigstar}$ & \cellcolor{green!25}\textbf{0.688}$^{\bigstar}$ & \cellcolor{green!25}\textbf{0.708}$^{\bigstar}$ & \cellcolor{green!25}\textbf{0.703}$^{\bigstar}$ & \cellcolor{green!25}\textbf{0.828}$^{\bigstar}$ & \cellcolor{green!25}\textbf{0.677}$^{\bigstar}$ & 0.543\\
MFCC + LogReg (n=40) & 80-dim & 69.0\% & 0.686 & 0.685 & 0.685 & 0.692 & 0.646 & 0.800 & 0.653 & \cellcolor{green!25}\textbf{0.642}$^{\bigstar}$\\
wav2vec2-base-superb-er & 768-dim & 43.3\% & 0.544 & 0.456 & 0.366 & 0.351 & \cellcolor{red!20}0.236$^{\dagger}$ & 0.618 & 0.136 & 0.474\\
wav2vec2-large-superb-er & 1024-dim & 48.8\% & 0.642 & 0.458 & 0.394 & 0.400 & 0.491 & 0.726 & 0.010 & 0.348\\
ESN + Ridge (n\_res=500) & 1000-dim & 68.3\% & 0.673 & 0.681 & 0.676 & 0.683 & \cellcolor{green!25}\textbf{0.703}$^{\bigstar}$ & 0.768 & 0.607 & 0.624\\
\bottomrule
\end{tabular}%
}
\end{table}

\begin{figure}[htbp]
\centering
\includegraphics[width=0.56\linewidth]{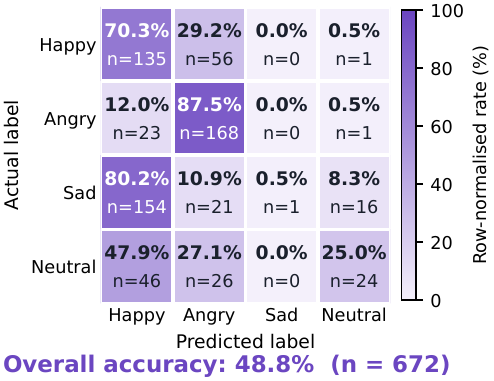}
\caption{Confusion matrix for wav2vec2-large-superb-er on RAVDESS (n\,=\,672), row-normalised
rates (\%) with per-cell counts. Sad recall collapses to 0.5\% under zero-shot transfer.}
\label{fig:confusion}
\end{figure}

The confusion matrix (Fig.~\ref{fig:confusion}) localises the failure: Angry recall reaches 87.5\% but Sad collapses to 0.5\%, most Sad clips landing as Happy or Angry, because RAVDESS's acted Sad style carries higher arousal than the IEMOCAP-fine-tuned model expects.
Both in-domain models recover Sad (MFCC F1\,=\,0.677, ESN 0.607 \cite{ibrahim2022}), confirming a transfer effect rather than an absence of Sad signal in the acoustics.

\paragraph{Emotion-grounded Q\&A utility (EQ2).}
Removing emotion evidence makes the model refuse most questions: 8/12 (67\%) of transcript-deployed answers are refusals, against 1/12 (8\%) when emotion evidence is present (Table~\ref{tab:qa}).
A refusal is the guardrail's templated reply that the question cannot be answered from the interview's emotional data; the model treats emotion-keyed questions as unanswerable without emotion input.
Softening Rule~6 leaves the refusal rate unchanged at 67\%, so the collapse reflects a genuine epistemic dependency rather than a citation-policy artefact.
On the four questions where all conditions answered substantively, every rubric mean falls between 3.69 and 4.38, with the three conditions within 0.26 of each other on each dimension and no ordering consistent across dimensions.
No reliable difference emerges (all $p\geq0.25$), and near-zero to negative Krippendorff's $\alpha$ ($-0.21$ to $0.11$) indicates poor rater agreement, so the comparison is inconclusive rather than a demonstrated tie.
Emotion grounding therefore \emph{enables} answers to emotion-keyed questions, not improving answers the model could already give from text alone.
The single-recording design makes EQ2 an exploratory utility test, not a definitive human evaluation.

\begin{table}[htbp]
\centering
\footnotesize
\caption{Q\&A annotation study (4 raters); conditions as in Table~\ref{tab:conditions}. Means and
SDs cover every non-refused answer per condition (11 questions emotion-informed, 4 each
transcript), so the columns are not on identical question sets. $\alpha$ is Krippendorff's
inter-rater reliability; $p_{dep}$ and $p_{soft}$ are Wilcoxon tests of emotion-informed against
each transcript condition over the four questions all answered, where the minimum attainable $p$
is 0.125. Highest mean per row in bold$^{\bigstar}$, lowest marked $^{\dagger}$, shading
supplementary: position only, since the columns are not comparable and no difference is reliable.
Refusal rates: 8\%, 67\%, 67\%.}
\label{tab:qa}
\resizebox{\textwidth}{!}{%
\begin{tabular}{lcccrr}
\toprule
\textbf{Dimension} & \textbf{Emotion-inf.} & \textbf{Transcript dep.} & \textbf{Transcript soft.} & \textbf{$\alpha$} & \textbf{$p_{dep}$ / $p_{soft}$}\\
\midrule
Traceability & 4.16 $\pm$ 0.91 & \cellcolor{green!25}\textbf{4.25}$^{\bigstar}$ $\pm$ 0.68 & \cellcolor{red!20}4.12$^{\dagger}$ $\pm$ 0.89 & $-0.21$ & 1.000 / 0.250\\
Faithfulness & \cellcolor{red!20}3.89$^{\dagger}$ $\pm$ 0.72 & 3.94 $\pm$ 0.77 & \cellcolor{green!25}\textbf{4.00}$^{\bigstar}$ $\pm$ 0.63 & $-0.05$ & 1.000 / 0.750\\
Timestamp accuracy & \cellcolor{red!20}4.22$^{\dagger}$ $\pm$ 1.14 & 4.31 $\pm$ 0.70 & \cellcolor{green!25}\textbf{4.38}$^{\bigstar}$ $\pm$ 0.72 & $-0.08$ & 1.000 / 0.625\\
Emotional specificity & \cellcolor{green!25}\textbf{3.89}$^{\bigstar}$ $\pm$ 0.99 & \cellcolor{red!20}3.69$^{\dagger}$ $\pm$ 0.70 & 3.75 $\pm$ 0.93 & 0.11 & 0.500 / 0.500\\
Groundedness & 4.14 $\pm$ 0.90 & \cellcolor{red!20}4.06$^{\dagger}$ $\pm$ 0.93 & \cellcolor{green!25}\textbf{4.19}$^{\bigstar}$ $\pm$ 0.91 & $-0.12$ & 0.500 / 1.000\\
\bottomrule
\end{tabular}}
\end{table}

Table~\ref{tab:ablation} reports the panel ablation.
Replacing the three-model panel with a single Gemma~3:4B verdict leaves refusal behaviour unchanged (0/36 answers in both conditions, identical across all three repeats) and citation density comparable: 8.6 against 10.3 cited timestamps, 7.2 against 10.8 emotion terms and 8.9 against 9.3 percentage citations per answer, all well inside the per-answer spread.
On these automatic metrics the panel is therefore not what enables emotion-grounded answering; the emotion evidence is.
The one consistent difference is discipline: single-model answers average 330 words against 181 for the full panel, longer on all twelve questions, drifting past the deployed prompt's two-to-four-sentence rule.
The shuffled control is the sharper result: with emotion distributions permuted across segments, the model still refuses nothing (0/36) and cites the corrupted scores slightly more densely than intact ones (9.8 against 8.9 percentage citations, 8.4 against 8.6 timestamps per answer).
Emotion grounding therefore governs whether the model answers, not whether the answer is right, and the reliability of every answer rests on the upstream classifier, consistent with the boundary drawn in Section~\ref{sec:discussion}.

Model diversity, tested separately, leaves no trace in the observable form of the answers: running all three roles on Gemma~3:4B alone, deployed prompts unchanged, matches the three-family panel over the same questions and repeats (0/36 refusals in both, 9.6 against 9.8 cited timestamps, 165 against 158 words, per-answer standard deviations above 100).
The choice is not thereby shown inert, since these metrics count citations and length whereas the rationale concerns whether one family's interpretive habits dominate the verdict; testing that needs human judgement over matched single- and multi-family verdicts.

\begin{table}[htbp]
\centering
\footnotesize
\caption{Panel ablation and corrupted-evidence control (12 questions $\times$ 3 repeats, emotion-informed prompt throughout, automatic scoring, no human re-rating). Citation and length columns are per-answer means over 36 answers. Highest value in each column is shown in bold$^{\bigstar}$ and the lowest is marked $^{\dagger}$, with cell shading supplementary; no column is better-is-higher, so a longer answer or a denser citation count is not a better one.}
\label{tab:ablation}
\resizebox{\textwidth}{!}{%
\begin{tabular}{lrrrrr}
\toprule
\textbf{Condition} & \textbf{Refusals} & \textbf{Timestamps} & \textbf{Emotion terms} & \textbf{\% citations} & \textbf{Words}\\
\midrule
Panel (deployed) & 0/36 & 8.6 & \cellcolor{red!20}7.2$^{\dagger}$ & \cellcolor{red!20}8.9$^{\dagger}$ & 181\\
Single model (Gemma only) & 0/36 & \cellcolor{green!25}\textbf{10.3}$^{\bigstar}$ & \cellcolor{green!25}\textbf{10.8}$^{\bigstar}$ & 9.3 & \cellcolor{green!25}\textbf{330}$^{\bigstar}$\\
Shuffled (control) & 0/36 & \cellcolor{red!20}8.4$^{\dagger}$ & 7.4 & \cellcolor{green!25}\textbf{9.8}$^{\bigstar}$ & \cellcolor{red!20}173$^{\dagger}$\\
\bottomrule
\end{tabular}%
}
\end{table}

\paragraph{Local deployment feasibility (EQ3).}
End-to-end CPU runtime over three runs ranged 142--165\,s (mean 157\,s), a \gls{rtf} of $\approx$1.33.
Whisper transcription dominates (mean 86\,s, $\approx$55\%), followed by pyannote diarisation (34\,s, $\approx$22\%), the three LLM calls (27\,s, $\approx$17\%) and the emotion classifier (9\,s, $\approx$6\%).
ASR, not LLM inference, is the latency bottleneck on CPU.
Every run made zero outbound network calls, making the pipeline suitable for offline review but not yet live analysis without streaming ASR.

\section{Discussion}
\label{sec:discussion}

\paragraph{What the Sad collapse means downstream.}
Near-zero Sad recall bounds how far the analysis can be trusted.
Genuinely sad speech is usually scored as Happy or Angry, so the evidence handed to the panel can carry the wrong sign, and the indicators inherit the distortion: Composure and Valence inflate whenever Sad probability mass is misassigned to Happy.
The panel cannot repair what it never sees, and the shuffled control (Table~\ref{tab:ablation}) confirms it answers as readily from corrupted as from intact evidence.
Two properties bound the damage: every assertion stays traceable to the segments behind it, so a reviewer who re-listens can catch a misread, and all outputs are advisory.
Sadness-dependent analyses such as distress screening therefore sit outside the current envelope, and the indicators remain qualitatively rather than experimentally validated, so in-domain adaptation and broader human-judgement validation are the key missing pieces.

\paragraph{Why local deployment matters, and what it reasonably requires.}
Privacy is the system's most measurable property: keeping biometric-adjacent audio on the device removes the dominant data-protection risk of cloud SER \cite{aloufi2021,backstrom2025}.
Reasonable local compute is modest: every EQ3 measurement comes from a consumer eight-core CPU with no GPU, and the two-phase design bounds peak memory to the largest single model rather than the sum of all six, so a mid-range desktop or laptop should suffice for offline review, though only the test machine has been measured.
The classical baselines are cheaper still and deploy locally at a fraction of the footprint, but a classifier alone only produces labels.
Aggregating affect over time, answering free-form questions and citing the segments behind each assertion are functions of the \gls{llm} layer, which accounts for 27\,s of the mean 157\,s runtime, and the value claim rests on that conversational layer rather than on classification throughput.

\paragraph{How emotion metadata changes LLM output.}
Prompt-level conditioning is a practical alternative to multimodal fine-tuning \cite{lin2024,thimonier2025}, but the shuffled control shows it buys answerability rather than correctness: traceability is only as reliable as the classifier.

\section{Ethics and Limitations}
\label{sec:ethics}

\paragraph{Privacy.}
Vocal recordings are personal data under UK GDPR Art.~4(1) and emotion scores may attract special-category treatment under Art.~9 \cite{icogdpr2021}.
Local-only processing removes outbound transfer but does not replace a \gls{dpia}: production deployment must still document lawful basis, consent, retention and erasure, and re-validate as the EU AI Act \cite{euaiact2024} evolves.

\paragraph{Affective-inference risk, terminology and bias.}
Affect inference is \emph{not} deception detection: Composure is reported as bounded emotional stability, open to inspection in the interface.
Bias is measured, not hypothetical: the Sad-class collapse (F1\,=\,0.010) is a concrete cross-corpus bias, and because RAVDESS is dominated by North-American actors, accent, age and gender bias remain unmeasured and must be validated before use \cite{sedenberg2017}.

\paragraph{Limitations.}
\label{sec:limits}
Evaluation uses an acted English corpus, so ecological validity for spontaneous field audio is unestablished.
Dual-use risk means EmotionAI is released as a post-hoc, advisory-not-diagnostic tool, and a clinician sign-off layer would be required before any mental-health application \cite{katirai2024}.
The four-class taxonomy omits surprise, contempt and sarcasm, and small (3--4\,B) models occasionally drift from the output schema.
The rubric study used four raters on a single 117.9\,s recording, restricted to the four questions answered in all conditions, and the ablations used automatic surface metrics only, so they bound what the panel changes rather than how good its answers are.

\section{Conclusion}
\label{sec:conc}

EmotionAI turns imperfect per-segment evidence into timestamp-grounded, citation-constrained analysis, entirely on the local host.
The deployed classifier transfers zero-shot well below an in-domain MFCC comparator, so the value is not SER accuracy but preserving data locality while making affective evidence inspectable, timestamped and usable by a local reasoning layer.
Future work prioritises in-domain fine-tuning to recover Sad recall, a broader Q\&A study, streaming ASR, and bias auditing across accent, age and gender.

\subsubsection*{Acknowledgements}
We would like to thank the Department of Computer Science at the Nottingham Trent University for supporting this research.

\bibliographystyle{splncs04}
\bibliography{references}

\end{document}